\documentclass[secnumarabic,aps,prl,reprint,groupedaddress]{revtex4-1}

\usepackage[english]{babel}
\usepackage{graphicx}
\usepackage{color}

\begin{document}
	
\newcommand{\unit}[1]{\:\mathrm{#1}}            
\newcommand{\To}{\mathrm{T_0}}
\newcommand{\Tp}{\mathrm{T_+}}
\newcommand{\Tm}{\mathrm{T_-}}
\newcommand{\EST}{E_{\mathrm{ST}}}
\newcommand{\Rp}{\mathrm{R_{+}}}
\newcommand{\Rm}{\mathrm{R_{-}}}
\newcommand{\Rpp}{\mathrm{R_{++}}}
\newcommand{\Rmm}{\mathrm{R_{--}}}
\newcommand{\ddensity}[2]{\rho_{#1\,#2,#1\,#2}} 
\newcommand{\ket}[1]{\left| #1 \right>} 
\newcommand{\bra}[1]{\left< #1 \right|} 

\title{Intrinsic and extrinsic contributions to spin scattering in Pt}
	
\author{Ryan Freeman$^1$}
\author{Andrei Zholud$^1$}
\author{Zhiling Dun $^2$}
\author{Haidong Zhou $^2$}
\author{Sergei Urazhdin$^1$}
	
\affiliation{$^1$Department of Physics, Emory University, Atlanta, GA, USA.}
\affiliation{$^2$Department of Physics and Astronomy, Knoxville, TN, USA.}

\begin{abstract}
	\textbf{
We utilize nanoscale spin valves with Pt spacer layers to characterize spin scattering in Pt. Analysis of the spin lifetime determined from our measurements indicates that the extrinsic Elliot-Yafet spin scattering is dominant at room temperature, while the intrinsic Dyakonov-Perel mechanism dominates at cryogenic temperatures. The significance of the latter is supported by the suppression of spin relaxation in Pt layers interfaced with a ferromagnet, likely caused by the competition between the effective exchange and spin-orbit fields.
}
\end{abstract}

\maketitle
	

The interplay between electron's motion and its spin due to the spin-orbit interaction (SOI) opens unprecedented opportunities for the control of both spin and orbital degrees of freedom~\cite{STT_Ralph,Spintronics_2001,Spintronics_Parkin,Spin_Calitronics_Review,New_Perspectives_on_Rashba}. For instance, the spin Hall effect (SHE) results in generation of pure spin current flowing transverse to charge current~\cite{DP_prediction_of_SHE}, enabling electronic control of static and dynamic states of magnetization in metallic and insulating nanomagnets~\cite{STT_from_SHE_in_PMA, SHE_in_Pt_YIG,Switching_Magnetic_Insulators}. Extensive recent studies of materials that exhibit large SOI, including Pt, Ta, W, topological insulators, and alloys such as CuBi, have focused on identifying the intrinsic and extrinsic mechanisms controlling SOI, and characterizing the relevant parameters including the spin-orbit scattering rates, the spin Hall angle, and the effective spin-orbit field~\cite{BuhrmanReview2012,Giant_SHA_Tantalum,SHA_in_Tungsten,CuBi,JaffresSML2014,Belashchenko_SML,STFMR_SHA_Pt,SO_field_Seebeck_InSb}. Another relevant parameter is the spin diffusion length $\lambda$, defined as the length scale over which the spin polarization relaxes away from the external source, which is determined mostly by the spin scattering due to SOI. It is also the length scale for spin current generation via the SHE, and is thus directly related to material's performance in spin-Hall applications.

Pt is one of the most extensively studied spin-orbit materials, thanks to the large SOI effects~\cite{BuhrmanReview2012,BassExpCritReview2007,SpinPumping_FMPt_bilayers}, relatively low resistivity that minimizes Joule heating and current shunting in heterostructures, and low reactivity. A variety of approaches have been utilized to determine the parameters relevant to SOI in Pt such as the spin Hall angle and $\lambda$ ~\cite{BuhrmanReview2012,BassExpCritReview2007,Hoffmann_SDL_by_spinpumping,CasanovaTdependenceSDL2015,MO_Detection_of_SHE_in_Pt_thin_films,JaffresSML2014}. Nevertheless, the values and the mechanisms controlling these parameters are still debated. In particular, the reported values of
the spin Hall angle in Pt range from $0.004$ to over $0.1$~\cite{BuhrmanReview2012,ParkinTransparency2015,JaffresSML2014}, and those of $\lambda$ range from less than $1$~nm to over $10$~nm~\cite{Hoffmann_SDL_by_spinpumping,CasanovaTdependenceSDL2015,BuhrmanReview2012,BassExpCritReview2007,Estimating_SDL_from_ISHE_and_spin_pumping,MO_Detection_of_SHE_in_Pt_thin_films,JaffresSML2014,Nguyen201430,0.77_2016}. Such a large spread of the reported characteristics makes it challenging to establish the dominant contributions to spin-orbit effects and the mechanisms controlling them.

One of the main difficulties in analyzing SOI is posed by the interplay between the interfacial and bulk effects. For instance, measurements of spin current generated by SHE are inevitably affected by the spin relaxation at the Pt interfaces, and by its generation via the interfacial Rashba effect ~\cite{InterfaceReview}. Indeed, the apparent spin Hall angle has been shown to depend on the transparency of the interfaces~\cite{ParkinTransparency2015}. Measurements of $\lambda$ based on the spin absorption efficiency~\cite{CasanovaTdependenceSDL2015} are similarly affected by the spin relaxation at the interfaces. Furthermore, the spin-orbit effects at interfaces with ferromagnets may be affected by the temperature-dependent contribution from the proximity-induced magnetism~\cite{Hoffman_reducedSHA_prox2015}.

One approach that can unambiguously separate the interfacial from the bulk contributions to spin relaxation is based on the giant magnetoresistance (GMR) in ferromagnet/normal metal/ferromagnet (F/N/F) spin valves, with the studied material inserted in the spacer N~\cite{GMR}. The value of $\lambda$ is directly determined from the dependence on the material thickness, while the contribution of the interfaces -- from the dependence on the number of inserted spacers. An important advantage of this technique for the quantitative analysis is that electrical current flows normal to the studied layer, and therefore electron transport is described by the bulk material parameters even for ultrathin films. In contrast, techniques based on the in-plane current flow require an elaborate analysis of  thickness-dependent resistivities and current shunting~\cite{Nguyen201430,0.77_2016}. Although the GMR-based approach to the quantitative characterization of spin scattering in materials is well established, only one such measurement has been reported for Pt at temperature $T=4.2$~K~\cite{Nguyen201430}, yielding  the value of $\lambda$ significantly larger than those reported based on other techniques~\cite{BassExpCritReview2007}.

Here, we report a study of the temperature-dependent GMR in nanopillar spin valves with Pt spacers, and demonstrate that this approach can be utilized to elucidate the mechanisms of spin scattering, and to characterize the relevant spin-orbit parameters. The dependence of GMR on the Pt spacer thickness allowed us to determine the value of $\lambda$ and extract the spin lifetime. The observed temperature dependence indicates that at room temperature, spin relaxation in the studied Pt films is dominated by extrinsic spin scattering, while at cryogenic temperatures it is dominated by spin dephasing due to the effective spin-orbit field. The flexibility of the GMR spin valve geometry also allowed us to show that the spin relaxation is suppressed in Pt at the interface with a ferromagnet. This observation is consistent with the competition between the effective spin-orbit field and the proximity-induced exchange field in Pt.

\begin{figure}
	\includegraphics[width=\columnwidth]{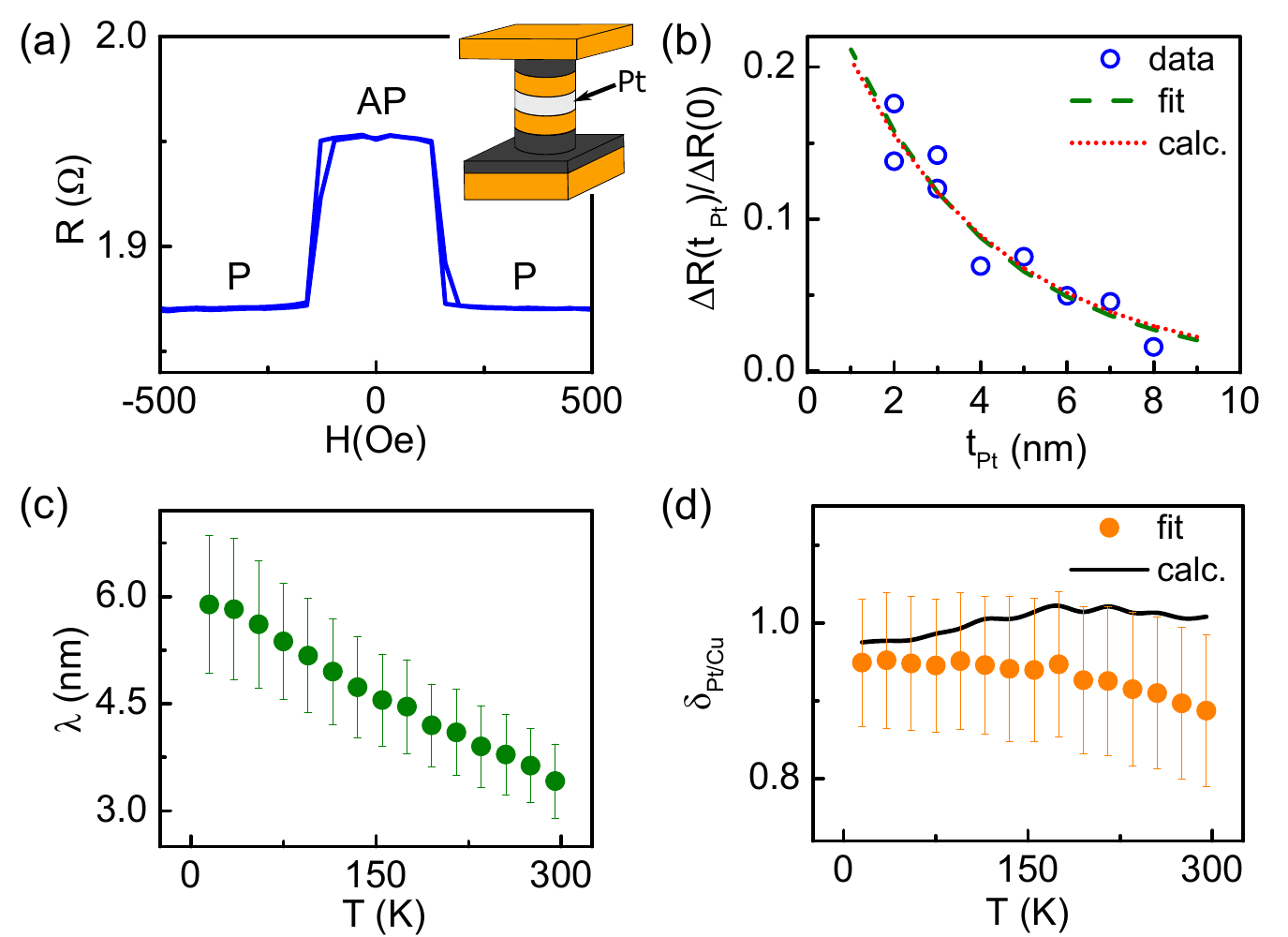}	
	\caption{\label{fig:Fig1} (Color online) (a) Resistance vs applied field for a sample without Pt spacer, at $RT=295$~K. Inset: schematic of the studied nanopillars, with Py layers shown in black, Cu in orange, and Pt in white. (b) Symbols: $\Delta R$ vs Pt thickness $d$, at RT, scaled by the MR of the reference sample without Pt spacer. Dashed curve: exponential fit to the data. Dotted curve: calculation based on the Valet-Fert theory. \textbf{(c)} Spin diffusion length vs temperature, determined by fitting $\Delta R(d)$ with Eq.~(\ref{exponent}). \textbf{(d)} Interfacial spin-loss factor vs temperature, obtained by fitting $\Delta R(d)$ with Eq.~(\ref{exponent}) (symbols), and by minimizing the difference between the MR data and the Valet-Fert calculations (curve).}
\end{figure}

The studied structures [inset in Fig.~\ref{fig:Fig1}(a)] were based on multilayers Cu(40)Py(10)Cu(6-d/2)Pt(d)Cu(6-d/2)Py(5)Au(5), where Py=Ni$_{80}$Fe$_{20}$, and thicknesses are given in nanometers. The thickness $d$ of the Pt insert was varied between $0$ and $8$~nm in $1$~nm increments, with $d=0$ representing the reference Py/Cu/Py nanopillar. The RMS roughness of the multilayers determined by atomic force microscopy was $0.3$~nm. We excluded the sample with $d=1$~nm from our analysis, because of the possible discontinuity of the $1$~nm-thick Pt layer. The thickness of the Cu spacers separating Pt from the magnetic Py layers was at least $2$~nm, to eliminate the effects of proximity-induced magnetism in Pt~\cite{TransportMagneticProximity,Sergei_ProxM_Pt2013}. A structure where Pt was directly interfaced with Py to analyze such effects is separately discussed below. We used a combination of e-beam lithography and Ar ion milling to pattern the Py(5) layer, the nonmagnetic Cu/Pt/Cu spacer, and $5$~nm of the bottom Py(10) into a circular $75$~nm nanopillar disk. The nanopillars were contacted by a Cu(80) top electrode, which was electrically isolated from the bottom electrode by a SiO$_2$(15) layer. Differential resistance was measured in a pseudo-four probe geometry by the lock-in detection technique, with an ac current of 100~$\mu$A rms at a frequency of $1.3$~kHz.

The magnetizations of the Py layers formed antiparallel (AP) configuration with resistance $R_{AP}$ at small field $H$, due to the antiferromagnetic coupling between the nanopatterned magnetic layers~\cite{AFcoupling}. Sufficiently large $H$ aligned both magnetizations into a parallel (P) configuration with resistance $R_{P}$, resulting in a well-defined switching between P and AP states in field scans [Fig.~\ref{fig:Fig1}(a)]. At a given temperature $T$, the dependence of magnetoresistance (MR) $\Delta R=R_{AP}-R_P$ on the thickness $d$ of the Pt layer was well-approximated by the exponential 
\begin{equation}\label{exponent}
\Delta R(d)=\Delta R(0)e^{-d/\lambda-2\delta_{Pt/Cu}},
\end{equation}
as shown by the dashed curve in Fig.~\ref{fig:Fig1}(b) for room temperature (RT), $T=295$~K. Here, $\delta_{Pt/Cu}$ is the phenomenological parameter describing spin loss at the Pt/Cu interface~\cite{Belashchenko_SML_Interfaces,ParkinTransparency2015,BuhrmanTransparency2015}. The temperature-dependent values of $\lambda$ and $\delta_{Pt/Cu}$ determined by fitting the dependence of MR on $d$ with Eq.~(\ref{exponent}) are shown by symbols in Figs.~\ref{fig:Fig1}(c) and (d), respectively. The former increased from $3.5$~nm at RT to $6.0$~nm at $7$~K [Fig.~\ref{fig:Fig1}(c)]. The slight increase of the latter from $0.89$ at RT to $0.95$ at $7$~K was within the fitting uncertainty. 

Although the value of $\delta_{Pt/Cu}$ is not central to the analysis of SOI presented below, we briefly discuss it here. This parameter is generally well-defined only for diffuse interfaces, $\delta=w/\lambda_I$, where $\lambda_I$ is the effective spin diffusion length in the interfacial region, and $w$ is its width~\cite{Belashchenko_SML}. To establish whether this interpretation is consistent with the value of $\delta_{Pt/Cu}$ obtained from fitting with Eq.~(\ref{exponent}), we performed calculations based on the Valet-Fert (VF) theory of GMR~\cite{VF_theory,Theory_of_bipolar_spin_switch,van_Son_Boundary_Resistance}, using the extracted $\lambda$ and the known spin-dependent transport properties of Py, Cu and their interfaces~\cite{BassExpCritReview2007,Bass1999274}, while adjusting $\delta_{Pt/Cu}$ to minimize the difference between our data and the calculated MR~\cite{supplementary}. The calculated dependence of MR on $d$ reproduced the exponential form  Eq.~(\ref{exponent}) [dotted curve in Fig.~\ref{fig:Fig1}(b)]. The calculated value of $\delta_{Pt/Cu}$ [curve in Fig.~\ref{fig:Fig1}(d)] is close to that obtained by simple exponential fitting. These results suggest that the spin relaxation rate at the Cu/Pt interface is approximately temperature-independent, with the value of $\delta_{Pt/Cu}$ consistent with the prior GMR measurements in macroscopic spin valves at $4.2$~K~\cite{Bass_SML_at_4.2K_2002}.
 
To elucidate the mechanisms of spin relaxation in Pt, we analyze the relationship between spin diffusion and orbital electron transport. If the spin relaxation is dominated by the extrinsic Elliot-Yafet (EY) mechanism, which originates from spin-flipping associated with momentum scattering in the presence of SOI, then the spin-flip time $\tau_{sf}$ is expected to be proportional to the momentum scattering time $\tau_{p}$~\cite{Zutic_and_Fabian_review_2004}. Another possible spin relaxation mechanism, Dyakonov-Perel (DP) relaxation, is the result of precession around the momentum-dependent spin-orbit field $H_{SO}$, and has been extensively discussed in the context of Rashba-Dresselhaus effects in materials with broken inversion symmetry~\cite{originalDPtheory,Zutic_and_Fabian_review_2004,InterfaceReview,Stiles_BoltzmannEqn_for_ST_including_Rashba2013,Boross_UnificationofDPandEY2013}. While Pt is inversion-symmetric, non-vanishing $H_{SO}$, known as the spin Berry curvature, is allowed by symmetry, and can contribute to both spin relaxation and the intrinsic SHE~\cite{HoffmannSHEreview,Guo_Intrinsic_SHE_Pt2008,Kontani_PRL_SHE_from_hybridization_2009,Berry_Curvature_Review,0.77_2016}. If the spin relaxation is dominated by the DP mechanism, then $\tau_{sf}$ should scale inversely with $\tau_{p}$.

\begin{figure}
	\centering
	\includegraphics[width=\columnwidth]{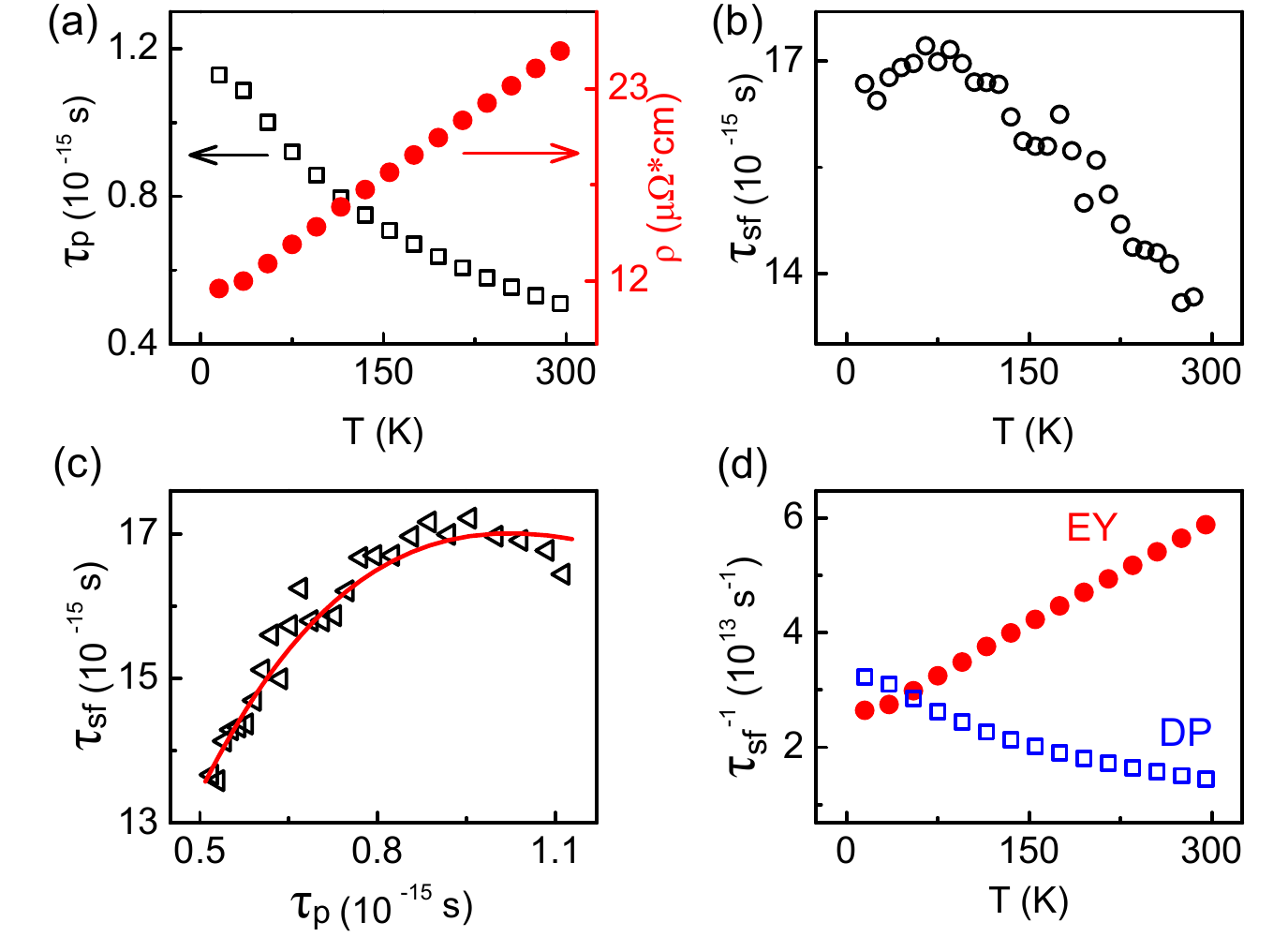}
	\caption{(Color online)(a) Bulk resistivity (right scale) and the momentum relaxation time (left scale) vs temperature for the studied Pt films. (b) Spin relaxation time vs temperature determined from the data of Fig.~\ref{fig:Fig1}(c). (c) Spin relaxation time vs momentum relaxation time. Symbols: data, curve: fitting with a superposition of EY and DP contributions, as described in the text. (d) Temperature dependence of the contributions to spin relaxation from DP and EY mechanisms, as labeled, determined from the fitting in panel (c).}
\label{fig:Fig2}
\end{figure}

We determined the momentum scattering time $\tau_p=m/\rho n e^2$ from the independently measured resistivity $\rho(T)$ [Fig.~\ref{fig:Fig2}(a)]~\cite{supplementary}, which is typical for sputtered Pt films~\cite{BassExpCritReview2007,BuhrmanReview2012}. Here, $m$ is the electron mass, $e$ is the electron charge, and $n=2.81\cdot 10^{29} m^{-3}$ is the carrier concentration determined by a separate Hall effect measurement~\cite{supplementary}. Figure~\ref{fig:Fig2}(b) shows the temperature dependence of the spin-flip time $\tau_{sf}$, determined from $\lambda$ [Fig.~\ref{fig:Fig1}(c)] using the spin diffusion relation $\lambda=\sqrt{v_Fl\tau_{sf}/3}$~\cite{BassExpCritReview2007}, where $v_F=\sqrt[3]{3\pi^2n}\hbar/m$ is the Fermi velocity, and $l=v_F\tau_p$ is the mean free path. At high temperatures, both $\tau_p$ and $\tau_{sf}$ linearly increase with decreasing temperature, indicating that the spin relaxation is dominated by the EY mechanism. While the dependence $\tau_p(T)$ remains monotonic at low temperatures, $\tau_{sf}$ starts to decrease, indicating a significant contribution of the DP  mechanism. To quantitatively analyze these contributions, we fit the relation between $\tau_{sf}$ and $\tau_p$ with
\begin{equation}\label{relaxation}
1/\tau_{sf}=1/\tau_{DP}+1/\tau_{EY},
\end{equation}
where $\tau_{DP}=1/(\Omega^2_{SO}\tau_p)$ is the DP contribution to spin relaxation in the spin random walk limit, and $\tau_{EY}=b^2\tau_p$ is the EY contribution~\cite{Zutic_and_Fabian_review_2004}. Here, $\Omega_{SO}=g\mu_BH_{SO}/\hbar$ is the average precession frequency around $H_{SO}$, and $b$ is the Eliott-Yafet parameter describing the band spin-mixing due to SOI, and related to the probability $P_{sf}=b^2/(1-b^2)$ of spin flipping per momentum scattering event~\cite{Boross_UnificationofDPandEY2013}. The fitting shown in Fig.~\ref{fig:Fig2}(c) slightly underestimates the decrease of $\tau_{sf}$ at large $\tau_p$, which may be caused by the enhancement of $H_{SO}$ at low temperatures neglected in our analysis~\cite{Hoffman_reducedSHA_prox2015}. The fitting allows us to estimate two relevant spin-orbit parameters: the average value of spin-orbit field $H_{SO}\approx 960\pm10$~Oe, and the band spin-mixing parameter $b\approx 0.17\pm0.02$, where the uncertainties reflect the accuracy of the fitting. The average precession phase $\Omega\cdot\tau_p\approx0.2$ between momentum scattering events satisfies the random spin-walk approximation used in our analysis, and the value of $b$ gives one spin flip per $25$ momentum scattering events, consistent with the diffusive limit assumed in our definition of $\tau_{sf}$. Figure~\ref{fig:Fig2}(d) shows the calculated temperature-dependent contributions of EY and DP mechanisms to spin relaxation in the studied Pt films. The EY contribution is about $4$ times larger than the DP contribution at RT. The EY contribution decreases with decreasing temperature, while the DP contribution increases, becoming larger than that of EY at temperatures below $50$~K. The DP contribution is expected to increase, while the EY contribution should decrease with increasing purity of Pt.

\begin{figure}
	\centering
	\includegraphics[width=\columnwidth]{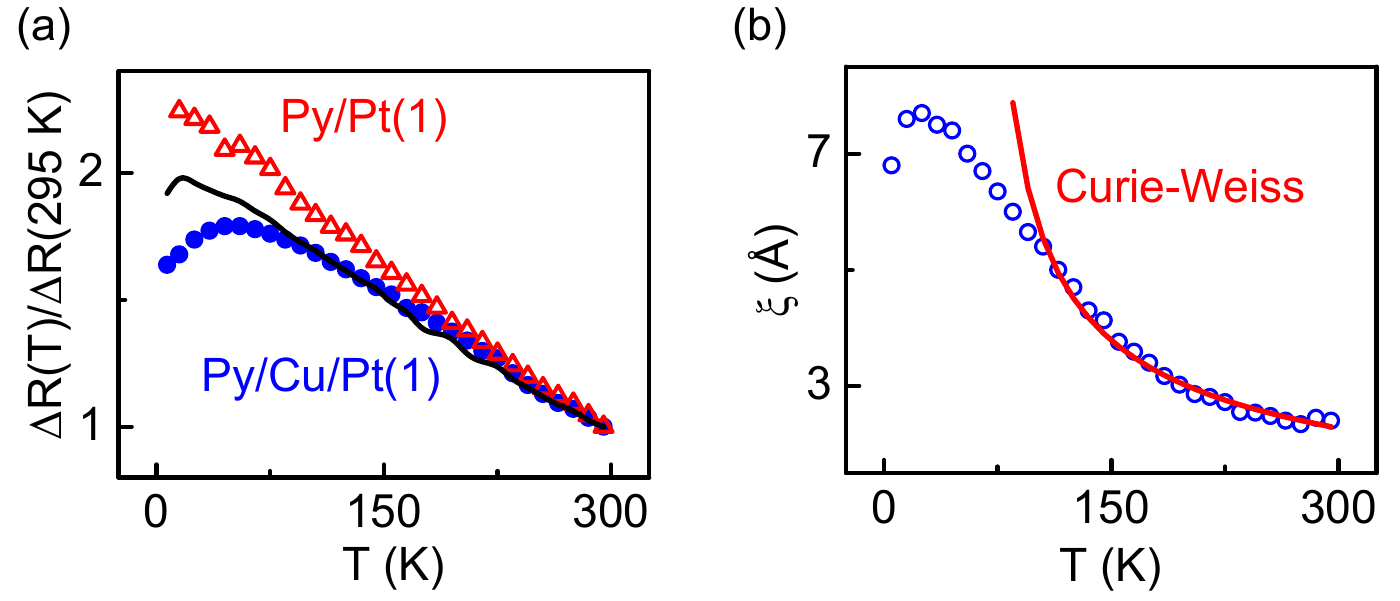}
	\caption{(a) MR vs temperature for the standard spin valve nanopillar with $d=1$~nm (solid symbols), and for the nanopillar with the structure Py(10)Pt(1)Cu(4)Py(5) (open symbols). Curve: VF calculation. The MR is normalized by its value at RT. \textbf{(d)} Symbols: magnetic correlation length in Pt vs temperature, from Ref.~\cite{Sergei_ProxM_Pt2013}. Curve: the dependence expected based on the Curie-Weiss law.}
	\label{fig:Fig3}
\end{figure}

The effect of the spin-orbit field on spin relaxation was supported by measurements of MR in separately fabricated nanopillars, where Pt was directly interfaced with one of the magnetic Py layers forming the spin valve. The difference with the standard structure discussed above was most prominent for the thinnest Pt layers with $d=1$~nm. For the standard structure, the MR linearly increases with decreasing $T>50$~K, and decreases at $T<50$~K [solid symbols in Fig.~\ref{fig:Fig3}(a)]. For thicker Pt, we observe a less pronounced curving at small T~\cite{supplementary}. The decrease of MR at low temperatures, associated with the increasing contribution of DP relaxation, is also evident in the VF calculations [curve in Fig.~\ref{fig:Fig3}(a)]. It is less significant than in the experimental data, likely due to the limitations of the diffusive transport model and/or the discontinuity of the Pt(1) layer. In contrast, for the nanopillar with the same geometry but active spin valve structure Py(10)Pt(1)Cu(4)Py(5), where Pt is directly interfaced with Py, the MR varies linearly with temperature [open symbols in Fig.~\ref{fig:Fig3}(a)].

The difference between the behaviors of the two structures can be correlated with the temperature dependence of the magnetic properties of Pt, as illustrated in Fig.~\ref{fig:Fig3}(b) reproduced from Ref.~\cite{Sergei_ProxM_Pt2013}. Symbols show the measured magnetic correlation length, and the curve shows the temperature dependence expected from the Curie-Weiss law, which  extrapolates to the Curie temperature of about $90$~K. However, the correlation length significantly deviates from the Curie-Weiss law at temperatures below $110$~K, and starts to decrease at $T<25$~K, suggesting that magnetism in Pt is increasingly suppressed at low temperatures. Indeed, bulk Pt remains an exchange-enhanced paramagnet even at cryogenic temperatures, while ferromagnetism was reported only in ultrathin films and nanoparticles~\cite{Emergent_Magnetism_in_Pt_nanowires,Synthesis_of_Pt_nanowires,FM_in_Pt_Ni_alloys}.

\begin{figure}
	\centering
	\includegraphics[width=\columnwidth]{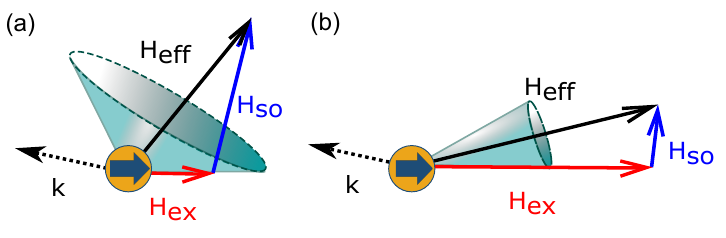}
	\caption{ Interplay between magnetism and intrinsic spin-orbit interaction. The spin of electron in Pt precesses around the effective field $\vec{H}_{eff}=\vec{H}_{SO}+\vec{H}_{ex}$, where $\vec{H}_{SO}$ is the effective spin-orbit field determined by the electron's momentum $k$, and $\vec{H}_{ex}$ is the effective exchange field due to either proximity-induced magnetism or magnetic fluctuation in Pt. The electron spin is shown by a bold circled arrow. (a) If $H_{SO}$ dominates, precessional dephasing suppresses the spin polarization and/or magnetization fluctuation. (b) If  $H_{ex}$ dominates, the precession angle for electron spins aligned with $\vec{H}_{ex}$ is small, suppressing the DP relaxation mechanism.}
	\label{fig:Fig4}
\end{figure}

We note that both the effects of the Pt/F interface on MR and the suppression of magnetism in Pt are observed at cryogenic temperatures, where the DP relaxation becomes increasingly significant [see Fig.~\ref{fig:Fig2}(d)]. We explain these effects by the interplay between $H_{SO}$ and the effective exchange field $H_{ex}$. If $H_{ex}$ is small or negligible away from the magnetic interfaces or in Pt interfaced with non-magnetic materials, the electron's spin experiences a large-angle precession around the total effective field $\vec{H}_{eff}=\vec{H}_{SO}+\vec{H}_{ex}$ dominated by $\vec{H}_{SO}$, Fig.~\ref{fig:Fig4}(a). Since the latter depends on the electron's wavevector $\vec{k}$, the spins of the diffusing electrons characterized by a broad distribution of $\vec{k}$ are efficiently dephased. This is the mechanism underling the DP relaxation. It is also likely responsible for the suppression of magnetism in Pt, since the precession, and consequently dephasing, of  electron spins comprising a magnetic fluctuation competes with the exchange interaction stabilizing such a fluctuation. This effect can be also described as quenching of spin polarization by SOI.

We now consider the situation where $H_{ex}$ is larger than $H_{SO}$, for example due to the proximity-induced magnetism in Pt near the interface with a ferromagnet. In this case, the spin precesses around the effective field $\vec{H}_{eff}$ dominated by $\vec{H}_{ex}$ [Fig.~\ref{fig:Fig4}(a)]. Since the spin of the electron injected from the ferromagnet is aligned with $\vec{H}_{ex}$, the precession angle is small, resulting in the suppression of the precessional spin relaxation, and thus an enhanced MR in the samples with proximity magnetized Pt. A similar mechanism has been proposed for the effect of external field on the DP relaxation~\cite{Dzhioev_Suppression_DP_2004}. This mechanism is also supported by the observed reduction of the spin Hall angle in proximity-magnetized Pt, likely caused by the suppression of the intrinsic SHE ~\cite{Hoffman_reducedSHA_prox2015}. 

The proposed mechanism of interplay between SOI and magnetism suggests that proximity magnetism in Pt should be enhanced when SOI-induced spin dephasing is reduced. Indeed, the coupling between two ferromagnets mediated by the proximity-magnetized Pt spacer was shown to monotonically increase with decreasing temperature~\cite{Sergei_ProxM_Pt2013}, even though the magnetic correlation length decreases at $T<25$~K [Fig.~\ref{fig:Fig3}b]. This is consistent with nonlinear enhancement of magnetism in Pt in the immediate vicinity of the interface, where DP relaxation is suppressed, compensating for its more abrupt decrease away from the interface.

In conclusion, we utilized nanoscale magnetic spin valves with Pt inserts to measure spin scattering in Pt, determine the temperature-dependent spin diffusion length, the bulk momentum and the spin relaxation times, and the spin relaxation at the interface with Cu. Analysis of the temperature dependence of these characteristics allowed us to separate the Elliott-Yafet from the Dyakonov-Perel contribution to spin relaxation, which result from the extrinsic scattering and the intrinsic spin dephasing, respectively. Our analysis allowed us to estimate the strength of the intrinsic spin-orbit field in Pt, and the parameter describing spin-subband mixing due to spin-orbit interaction. Furthermore, we show that spin-orbit effects can be affected by the proximity-induced magnetism at the interfaces of Pt with ferromagnets, resulting in suppression of spin relaxation. Our results demonstrate an efficient route for the quantitative characterization of spin-orbit interactions, which should facilitate the exploration and design of new efficient spin-orbit materials.

This work was supported by the NSF grant Nos. DMR-1350002 and DMR-1504449. 

\bibliography{SDLbib}
\bibliographystyle{unsrt}

\end{document}